\documentclass[12pt]{article}

\usepackage[margin=1in,nohead,dvips,a4paper]{geometry}
\usepackage{times}
\usepackage{amsfonts,amssymb}

\makeatletter
\def\eqnarray{%
   \stepcounter{equation}%
   \def\@currentlabel{\p@equation\theequation}%
   \global\@eqnswtrue 
   \m@th
   \global\@eqcnt\z@
   \tabskip\@centering
   \let\\\@eqncr
   $$\everycr{}\halign to\displaywidth\bgroup
       \hskip\@centering$\displaystyle\tabskip\z@skip{##}$\@eqnsel
      &\global\@eqcnt\@ne\hfil$\displaystyle{\hbox{}##\hbox{}}$\hfil
      &\global\@eqcnt\tw@ $\displaystyle{##}$\hfil\tabskip\@centering
      &\global\@eqcnt\thr@@ \hb@xt@\z@\bgroup\hss##\egroup
         \tabskip\z@skip
      \cr
}
\def\lefteqn#1{\hbox to 2em{$\displaystyle #1$\hss}} 
\DeclareRobustCommand{\qed}{%
  \ifmmode 
  \else \leavevmode\unskip\penalty9999 \hbox{}\nobreak\hfill
  \fi
  \quad\hbox{\qedsymbol}}
\newcommand{\openbox}{\leavevmode
  \hbox to.77778em{%
  \hfil\vrule
  \vbox to.675em{\hrule width.6em\vfil\hrule}%
  \vrule\hfil}}
\newcommand{\qedsymbol}{\openbox}
\newenvironment{proof}[1][\proofname]{\par
  \normalfont
  \topsep6\p@\@plus6\p@ \trivlist
  \item[\hskip\labelsep\itshape
    #1.]\ignorespaces
}{%
  \qed\endtrivlist
}
\newcommand{\proofname}{Proof}
\makeatother
\def\Ad {\mathop{\mathrm{Ad}}}
\def\ad {\mathop{\mathrm{ad}}}
\def\Aut {\mathop{\mathrm{Aut}}}
\def\bbC {\mathbb C}
\def\bbN {\mathbb N}
\def\bbR {\mathbb R}
\def\bbZ {\mathbb Z}

\def\calD {\mathcal D}

\def\calL {\mathcal L}

\def\Der {\mathop{\mathrm{Der}}}
\def\Diff {\mathop{\mathrm{Diff}}}
\def\gothG {\mathfrak G}
\def\gothg {\mathfrak g}
\def\gothH {\mathfrak H}

\def\id {\mathrm {id}}
\def\Int {\mathop{\mathrm{Int}}}
\def\rmd {\mathrm d}
\def\rme {\mathrm e}
\def\rmi {\mathrm i}
\def\supp {\mathop{\mathrm{supp}}}

\mathchardef\leq="3436
\mathchardef\geq="343E
\mathchardef\slantLambda="103
\mathchardef\slantPhi="108
\mathchardef\slantPi="105
\mathchardef\slantPsi="109
\mathchardef\slantTheta="102

\DeclareRobustCommand{\,}{%
   \relax\ifmmode\mskip .5\thinmuskip\else\thinspace\fi
}

\def\mbar#1{\kern 0.1em 
 \overline{\kern -0.1em #1 \kern -0.1em} 
 \kern 0.1em}

\newtheorem{theorem}{Theorem}[section]
\newtheorem{proposition}{Proposition}[section]
\newtheorem{lemma}{Lemma}[section]

\title{On $\bbZ$-gradations of twisted loop Lie algebras\\
of complex simple Lie algebras}
\author{Kh. S. Nirov\thanks{On leave of absence from the Institute for
Nuclear Research of the Russian Academy of Sciences, 117312 Moscow, Russia.
E-mail: nirov@ms2.inr.ac.ru}\\
\small Max-Planck-Institut f\"ur Gravitationsphysik -- 
Albert-Einstein-Institut\\[-.3em]
\small Am M\"uhlenberg 1, D-14476 Golm b. Potsdam, Germany\\[-.3em]
\small E-mail: nirov@aei.mpg.de\\[.3em]
A. V. Razumov\\
\small Institute for High Energy Physics\\[-.3em]
\small 142280 Protvino, Moscow Region, Russia\\[-.3em]
\small E-mail: Alexander.Razumov@ihep.ru}
\date{}

\begin{document}

\maketitle

\begin{abstract}
We define the twisted loop Lie algebra of a finite dimensional Lie algebra
$\gothg$ as the Fr\'echet space of all twisted periodic smooth mappings
from $\bbR$ to $\gothg$. Here the Lie algebra operation is continuous. We
call such Lie algebras Fr\'echet Lie algebras. 
We introduce the notion of an integrable $\bbZ$-gradation of a Fr\'echet
Lie algebra, and find all inequivalent integrable $\bbZ$-gradations with
finite dimensional grading subspaces of twisted loop Lie algebras of
complex simple Lie algebras.
\end{abstract}

\section{Introduction}

The theory of loop groups and loop Lie algebras has a lot of applications
to mathematical and physical problems. In particular, it is a necessary
tool for formulation of many integrable systems and construction of
appropriate integration methods. Here one or another version of
factorization problem for the underlying group arises (see, for example,
\cite{Sem03}). For the so-called Toda systems associated with loop groups
the required factorization is induced by a $\bbZ$-gradation of the
corresponding loop Lie algebra, and, at least from this point of view, the
classification of $\bbZ$-gradations of loop Lie algebras is quite
important. The definition and general integration procedure for the Toda
systems can be found in \cite{LezSav92, RazSav97a, RazSav97b}. The
classification of $\bbZ$-gradations of complex semisimple finite
dimensional Lie algebras is well known (see, for example,
\cite{GorOniVin94}). The corresponding classification of the Toda systems
associated with compex classical Lie groups was given in papers
\cite{RazSavZue99, NirRaz03}.

There are two main definitions of the loop Lie algebras. In accordance with
the first definition used, for example, by Kac in his famous monograph
\cite{Kac94}, a loop Lie algebra is the set of finite Laurent polynomials
with coefficients in a finite dimensional Lie algebra. It is rather
difficult to associate a Lie group with such a Lie algebra. Actually this
is connected to the fact that the exponential of a finite polynomial is
usually not a finite polynomial. However, it should be noted that with
this definition in the case when the underlying Lie algebra is complex and
simple one can classify all $\bbZ$-gradations of the loop Lie algebras with
finite dimensional grading subspaces~\cite{Kac94}\footnote{Actually in
\cite{Kac94} one can find the classification of $\bbZ$-gradations of the
affine Kac--Moody algebras. The classification of $\bbZ$-gradations of the
corresponding loop Lie algebras immediately follows from that
classification.}.

In accordance with the second definition, used in the monograph by Pressley
and Segal \cite{PreSea86}, a loop Lie algebra is the set of smooth mappings
from the circle $S^1$ to a finite dimensional Lie algebra. This set is
endowed with the structure of a Fr\'echet space. Here the Lie algebra
operation defined pointwise is continuous. The definition given in
\cite{PreSea86} is more convenient for applications to the theory of
integrable systems, because in this case we always have an appropriate Lie
group. Therefore, it would be interesting and useful to obtain a
classification of $\bbZ$-gradations for loop Lie algebras defined as in
\cite{PreSea86}. In the present paper we introduce the concept of an
integrable $\bbZ$-gradation and classify all integrable $\bbZ$-gradations
with finite dimensional grading subspaces of loop Lie algebras and twisted
loop Lie algebras of finite dimensional complex simple Lie algebras. The
result of the classification is actually the same as for loop Lie algebras
and twisted loop Lie algebras defined as in \cite{Kac94}. Namely, to
classify all integrable $\bbZ$-gradations with finite dimensional grading
subspaces of the Lie algebras under consideration one has to classify all
$\bbZ_K$-gradations of the underlying Lie algebras or, equivalently, all
their automorphisms of finite order.

\section{Loop Lie algebras and loop Lie groups} \label{s:2}

Consider the vector space $C^\infty(S^1, V)$ of smooth mappings from the
circle $S^1$ to a finite dimensional vector space $V$. It is convenient to
treat the circle $S^1$ as the set of complex numbers of modulus one. There
is a natural mapping from the set $\bbR$ of real numbers to $S^1$ which
takes $\sigma \in \bbR$ to $\rme^{\rmi \sigma} \in S^1$. Given an element
$\xi \in C^\infty(S^1, V)$, one defines a mapping $\tilde \xi$ from
$\bbR$ to $V$ by the equality
\[
\tilde \xi(\sigma) = \xi(\rme^{\rmi \sigma}).
\]
The mapping $\tilde \xi$ is smooth and satisfies the relation $\tilde
\xi(\sigma + 2 \pi) = \tilde \xi(\sigma)$. Conversely, any smooth
periodic mapping from $\bbR$ to $V$ induces a smooth mapping from
$S^1$ to $V$. Introduce the notation
\[
\tilde \xi^{(k)} = \rmd^k \tilde \xi/\rmd s^k,
\]
where $s$ is the standard coordinate function on $\bbR$. It is customary to
assume that $\tilde \xi^{(0)} = \tilde \xi$. Given an element $\xi \in
C^\infty(S^1, V)$, we denote by $\xi^{(k)}$ the element of $C^\infty(S^1,
V)$ induced by $\tilde \xi^{(k)}$.

Endow $C^\infty(S^1, V)$ with the structure of a topological vector space
in the following way. Let $\| \cdot \|$ be a norm on $V$. Define a
countable collection of norms $\{\| \cdot \|_m\}_{m \in \bbN}$ on
$C^\infty(S^1, V)$ by
\[
\|\xi\|_m = \max_{0 \leq k < m} \, \, \max_{p \in S^1} \, \, \|
\xi^{(k)}(p) \|,
\]
or via the corresponding mapping $\tilde \xi: \bbR \to V$ by
\[
\|\tilde \xi\|_m = \max_{0 \leq k < m} \, \, \max_{\sigma \in [0, 2 \pi]}
\, \, \| \tilde \xi^{(k)}(\sigma) \|.
\]
Note that for any $\xi \in C^\infty(S^1, V)$, if $m_1 < m_2$, then
$\|\xi\|_{m_1} \leq \|\xi\|_{m_2}$.

Given a positive integer $m$, denote
\[
U_m = \{\xi \in C^\infty(S^1, V) \mid \|\xi\|_m < 1/m\}.
\]
The collection formed by the sets $U_m$ is a local base of a topology on
$C^\infty(S^1, V)$. As a base of the topology we can take the collection of
subsets of the form
\[
U_{\xi, m} = \xi + U_m, \qquad \xi \in C^\infty(S^1, V).
\]
A sequence $(\xi_n)$ in $C^\infty(S^1, V)$ converges to $\xi \in
C^\infty(S^1, V)$ relative to this topology, if and only if for each
nonnegative integer $k$ the sequence $\left( \tilde \xi{}^{(k)}_n
\right)$ converges uniformly to $\tilde \xi{}^{(k)}$. One can show that
actually we have a Fr\'echet space. We define a Fr\'echet space as a
complete topological vector space whose topology is induced by a countable
family of semi-norms.\footnote{Sometimes a more general definition of a
Fr\'echet space is used (see, for example, \cite{Rud73}).}

Let now $\gothg$ be a real or complex finite dimensional Lie algebra.
Supply the Fr\'echet space $C^\infty(S^1, \gothg)$ with the Lie algebra
structure defining the Lie algebra operation pointwise. The obtained Lie
algebra is called the {\em loop Lie algebra\/} of $\gothg$ and denoted
$\calL (\gothg)$. It is clear that constant mappings form a subalgebra
of $\calL(\gothg)$ which is isomorphic to the initial Lie algebra
$\gothg$.

Let again $V$ be a finite dimensional vector space, and let $a$ be an
automorphism of $V$. Consider the quotient space $E$ of the direct product
$\bbR \times V$ by the equivalence relation which identifies $(\sigma, v)$
with $(\sigma + 2 \pi, a(v))$. Define the projection $\pi: E \to S^1$ by
the relation 
\[
\pi([(\sigma, v)]) = \rme^{\rmi \sigma}.
\]
It is not difficult to show that in such a way one obtains a smooth vector
bundle $E \stackrel{\pi\;}{\to} S^1$ with fiber~$V$.

Let $\xi$ be a smooth section of $E$. For any $\sigma \in \bbR$ there
exists a unique element $\tilde \xi(\sigma) \in V$ such that
\[
[(\sigma, \tilde \xi(\sigma))] = \xi(\rme^{\rmi \sigma}).
\]
This relation defines a smooth mapping $\tilde \xi$ from $\bbR$ to $V$
which satisfies the relation
\[
\tilde \xi (\sigma + 2 \pi) = a (\tilde \xi(\sigma)),
\]
called {\em twisted periodicity\/}. Conversely, given a mapping $\tilde
\xi: \bbR \to V$ which is twisted periodic, the equality
\[
\xi(p) = [(\sigma, \tilde \xi(\sigma))], \qquad p = \rme^{\rmi \sigma},
\]
defines a smooth section of $E$. One can make the space $C^\infty(S^1
\stackrel{\;\;\pi}{\leftarrow}E)$ of smooth sections of $E
\stackrel{\pi\;}{\to} S^1$ a Fr\'echet space in the same way as it was done
above for the space $C^\infty(S^1, V)$. Here it is natural and useful to
assume that the corresponding norm on $V$ is invariant with respect to the
automorphism~$a$.

If the vector space $V$ is a Lie algebra $\gothg$ and $a$ is an
automorphism of $\gothg$, one can supply the vector space $C^\infty(S^1
\stackrel{\;\;\pi}{\leftarrow}E)$, or equivalently the vector space of
twisted periodic mappings from $\bbR$ to $\gothg$, with the structure of 
a Lie algebra defining Lie algebra operation pointwise. We denote this Lie
algebra by $\calL_a(\gothg)$ and call a {\em twisted loop Lie algebra\/}. 
The loop Lie algebra $\calL(\gothg)$ can be considered as the twisted
loop Lie algebra $\calL_a(\gothg)$ with $a = \id_{\gothg}$.

Let $G$ be a Lie group whose Lie algebra coincides with $\gothg$ and $\Ad$
be the adjoint representation of $G$ in $\gothg$. For any $g \in G$ the
linear operator $\Ad(g)$ is an automorphism of $\gothg$. Such automorphisms
are called {\em inner automorphisms\/}. They form a normal subgroup $\Int
\gothg$ of the group $\Aut \gothg$ of automorphisms of $\gothg$. One can
show that if the automorphisms $a$ and $b$ of $\gothg$ differ by an inner
automorphism of $\gothg$ then the twisted loop Lie algebras
$\calL_a(\gothg)$ and $\calL_b(\gothg)$ are naturally isomorphic. This
means, in particular, that if the Lie algebra $\gothg$ is semisimple one
can consider only the twisted loop Lie algebras $\calL_a(\gothg)$ with $a$
belonging to the finite subgroup of $\Aut \gothg$ identified with the
automorphism group $\Aut \slantPi$ of some simple root system $\slantPi$ of
$\gothg$. In particular, one can assume that $a^K = \id_\gothg$ for some
positive integer $K$. It is convenient for our purposes to assume that $K$
does not necessarily coincide with the order of $a$.

Let $\gothg$ be a semisimple Lie algebra. Consider an arbitrary element
$\eta$ of $\calL_a(\gothg)$ and the corresponding mapping $\tilde \eta$
from $\bbR$ to $\gothg$. It is clear that the mapping $\tilde \xi$ defined
as
\[
\tilde \xi(\sigma) = \tilde \eta(K \sigma),
\]
is a periodic mapping from $\bbR$ to $\gothg$. Therefore, it induces an
element $\xi$ of $\calL(\gothg)$. It is clear that in this way we obtain an
injective homomorphism from $\calL_a(\gothg)$ to $\calL(\gothg)$. 
The image of this homomorphism is formed by the elements $\xi$ satisfying
the condition
\[
\xi(\varepsilon_K p) = a (\xi(p)),
\]
where $\varepsilon_K = \exp(2 \pi \rmi/K)$ is the $K$th principal root of
unity. We will denote this image as $\calL_{a,K}(\gothg)$. For the
corresponding mapping $\tilde \xi$ from $\bbR$ to $\gothg$
one has
\[
\tilde \xi(\sigma + 2 \pi/K) = a (\tilde \xi(\sigma)).
\]
Thus, when $\gothg$ is a semisimple Lie algebra, the twisted loop Lie
algebra $\calL_a(\gothg)$ can be identified with a subalgebra of the loop
Lie algebra $\calL(\gothg)$.

We call a Lie algebra $\gothG$ a {\em Fr\'echet Lie algebra\/} if $\gothG$
is a Fr\'echet space and the Lie algebra operation in $\gothG$, considered
as a mapping from $\gothG \times \gothG$ to $\gothG$, is continuous.
Actually one can consider a Fr\'echet Lie algebra as a smooth manifold
modelled on itself. Here the Lie algebra operation is a smooth mapping.

To prove that $\calL_a(\gothg)$ is a Fr\'echet Lie algebra we start with
the following simple lemmas.

\begin{lemma} \label{l:2}
Let $\gothg$ be a finite dimensional Lie algebra and $\|\cdot\|$ be a
norm
on $\gothg$. There exists a positive real number $C$ such that
\[
\|[x, y]\| \leq C \, \|x\| \, \|y\|
\]
for all $x, y \in \gothg$.
\end{lemma}
\begin{proof} 
First prove the statement of the proposition for a special choice of
the
norm $\|\cdot\|$. Let $(e_i)$ be a basis of $\gothg$. Expand an
arbitrary
element $x$ of $\gothg$ over the basis $(e_i)$, $x = \sum_i e_i x^i$,
and
define
\[
\|x\| = \max_i \{|x^i|\}.
\]
In this case for any $i$ one has $|x^i| \leq \|x\|$. It is also
evident that $\|e_i\| = 1$. For arbitrary elements $x, y \in \gothg$
one
has
\[
[x, y] = \sum_{i,j}[e_i x^i, e_j y^j] = \sum_{i,j,k} e_k \, c^k{}_{ij}
\,
x^i y^j,
\]
where $c^k{}_{ij}$ are structure constants of the Lie algebra
$\gothg$,
\[
[e_i, e_j] = \sum_k e_k \, c^k{}_{ij}.
\]
Therefore,
\[
\|[x, y]\| \leq \sum_{i,j,k} \|e_k\| \, |c^k{}_{ij}| \, |x^i| \, |y^j|
\leq \left( \sum_{i,j,k} |c^k{}_{ij}| \right) \|x\| \|y\|.
\]
Thus, the statement of the proposition is valid for the norm chosen
and for
\[
C = \sum_{i,j,k} |c^k{}_{ij}|.
\]
Since in the finite dimensional case all norms are equivalent, the
statement of the proposition is valid for an arbitrary norm
$\|\cdot\|$.
\end{proof}

\begin{lemma} \label{l:1}
There are positive real numbers $C_m$, $m = 1, 2, \ldots$, such that
\[
\|[\xi, \eta]\|_m \leq C_m \, \|\xi\|_m \, \|\eta\|_m
\]
for all $\xi, \eta \in \calL_a(\gothg)$.
\end{lemma}

\begin{proof}
For $m=1$, using Lemma \ref{l:2}, one easily obtains
\begin{eqnarray*}
\|[\xi, \eta]\|_1 &=& \max_{p \in S^1} \| [\xi(p), \eta(p)] \| \leq \max_{p
\in S^1} C \, \| \xi(p)\| \, \| \eta(p) \| \leq C \, \| \xi \|_1 \, \| \eta
\|_1.
\end{eqnarray*}
For $m=2$ one has
\[
\|[\xi, \eta]\|_2 = \max \left\{ \max_{p \in S^1} \| [\xi(p), \eta(p)] \| ,
\, \max_{p \in S^1} \| [\xi, \eta]^{(1)}(p) \| \right\}.
\]
It is clear that
\begin{eqnarray*} 
\max_{p \in S^1} \| [\xi, \eta]^{(1)}(p) \| &&  = \max_{p \in S^1} \|
[\xi^{(1)}, \eta](p) + [\xi, \eta^{(1)}](p) \| \\
&& \leq C \, \| \xi^{(1)} \|_1 \, \|\eta\|_1 + C \, \| \xi\|_1 \, \|
\eta^{(1)} \|_1 \leq 2 \, C \, \|\xi\|_2 \, \|\eta\|_2.
\end{eqnarray*}
Taking into account that $\| \cdot \|_1 \leq \| \cdot \|_2$, we conclude
that
\[
\|[\xi, \eta]\|_2 \leq 2 \, C \|\eta\|_2 \, \|\xi\|_2.
\]
Similarly one obtains
\[
\|[\xi, \eta]\|_m \leq 2^{m-1} \, C \|\xi\|_m \, \|\eta\|_m.
\]
Thus, the statement of the proposition is valid for $C_m = 2^{m-1} C$.
\end{proof}
Now we are able to prove the desired result.
\begin{proposition}
The twisted loop Lie algebra $\calL_a(\gothg)$ is a Fr\'echet Lie algebra.
\end{proposition}

\begin{proof}
It suffices to show that for any fixed elements $\xi_1, \xi_2 \in
\calL_a(\gothg)$ and any positive integer $m$, there are positive integers
$m_1$ and $m_2$ such that for any $\xi'_1 \in U_{\xi_1, m_1}$ and $\xi'_2
\in U_{\xi_2, m_2}$ one has $[\xi'_1, \xi'_2] \in U_{[\xi_1, \xi_2],m}$. It
is clear that one can assume that $m_1 \geq m$ and $m_2 \geq m$. 

Let $m$ be a fixed positive integer, $\xi_1$ and $\xi_2$ be arbitrary
elements of $\calL(\gothg)$, $m_1$, $m_2$ be arbitrary positive integers
greater than $m$. For any $\xi'_1 \in U_{\xi_1,m_1}$ and $\xi'_2 \in
U_{\xi_2, m_2}$ write the equalities
\begin{eqnarray*}
[\xi'_1, \xi'_2] - [\xi_1, \xi_2] &=& [(\xi'_1 - \xi_1) + \xi_1, (\xi'_2 -
\xi_2) + \xi_2] - [\xi_1, \xi_2] \\[.5em]
&=& [\xi'_1 - \xi_1, \xi'_2 - \xi_2] + [\xi_1, \xi'_2 - \xi_2] +
[\xi'_1 - \xi_1, \xi_2].
\end{eqnarray*}
Using Lemma \ref{l:1}, we obtain
\begin{eqnarray*}
\| [\xi'_1, \xi'_2] - [\xi_1, \xi_2] \|_m &\leq& C_m (\| \xi'_1 - \xi_1
\|_m \| \xi'_2 - \xi_2 \|_m \\[.5em]
&& \hspace{2em} {} + \| \xi'_1 - \xi_1 \|_m \|\xi_2 \|_m  + \|\xi_1 \|_m \|
\xi'_2 - \xi_2 \|_m ) \\[.5em]
&\leq& C_m (\| \xi'_1 - \xi_1 \|_{m_1} \| \xi'_2 - \xi_2 \|_{m_2} \\[.5em]
&& \hspace{2em} {} + \| \xi'_1 - \xi_1 \|_{m_1} \| \xi_2 \|_m  + \| \xi_1
\|_m \| \xi'_2 - \xi_2 \|_{m_2} ).
\end{eqnarray*}
Thus, we have
\[
\| [\xi'_1, \xi'_2] - [\xi_1, \xi_2] \|_m < C_m \left( \frac{1}{m_1}
\frac{1}{m_2} + \frac{1}{m_1} \| \xi_2 \|_m + \| \xi_1 \|_m
\frac{1}{m_2}
\right).
\]
It is clear that for sufficiently large $m_1$ and $m_2$ one has
\[
\| [\xi'_1, \xi'_2] - [\xi_1, \xi_2] \|_m < 1/m,
\]
that means that $[\xi'_1, \xi'_2] \in U_{[\xi_1, \xi_2],m}$.
\end{proof}

Let $G$ be a finite dimensional Lie group with the Lie algebra $\gothg$.
The loop group $\calL(G)$ is defined as the set of all smooth mappings
from the circle $S^1$ to $G$ with the group law being pointwise composition
in $G$. Here, as for the case of loop Lie algebras, for any element
$\gamma$ of $\calL(G)$ one can define a smooth mapping $\tilde \gamma$ 
from $\bbR$ to $G$ connected with $\gamma$ by the equality
\[
\tilde \gamma(\sigma) = \gamma(\rme^{\rmi \sigma}),
\]
and satisfying the relation $\tilde \gamma(\sigma + 2 \pi) = \tilde
\gamma(\sigma)$. Conversly, any periodic smooth mapping from $\bbR$
to $G$ induces an element of $\calL(G)$.

One can endow the loop group $\calL(G)$ with the structure of an infinite
dimensional manifold and a Lie group in the following way.

Recall that the exponential mapping $\exp : \gothg \to G$ is a local
diffeomorphism near the identity. Let $\breve U_e$ be an open neighbourhood
of the identity of $G$ diffeomorphic to some open neighbourhood of the zero
element of $\gothg$, and $\breve \varphi$ be the restriction of the inverse
of the exponential mapping to $\breve U_e$. Denote 
$U_e = C^\infty(S^1, \breve U_e)$ and define a mapping 
$\varphi: U_e \to C^\infty(S^1, \breve \varphi(\breve U_e))$ by
\[
\varphi(\gamma) = \breve \varphi \circ \gamma.
\]
Note that the set $C^\infty(S^1, \breve \varphi(\breve U_e))$ is open in
$\calL(\gothg)$ and we can consider the pair $(U_e, \varphi)$ as a chart on
$\calL(G)$. 

For an arbitrary element $\gamma \in \calL(G)$ denote $U_\gamma =
\gamma \, U_e$, and define the mapping $\varphi_\gamma: U_\gamma \to
C^\infty(S^1, \breve \varphi(\breve U_e))$ by
\[
\varphi_\gamma(\gamma') = \breve \varphi \circ (\gamma^{-1} \gamma').
\]
In this way we obtain an atlas which makes $\calL(G)$ into a smooth
manifold modelled on the Fr\'echet space $\calL(\gothg)$. Actually 
in this way $\calL(G)$ becomes a Lie group. The Lie algebra of
$\calL(G)$
can be naturally identified with $\calL(\gothg)$.

We say that the set $U \subset \calL(G)$ is open if for any $\gamma \in
\calL(G)$ the set $\varphi_\gamma(U \cap U_\gamma)$ is open. This
definition supplies $\calL(G)$ with the structure of a topological
space. As any Lie group the loop Lie group $\calL(G)$ is a Hausdorff
topological space.

Twisted loop groups are defined in full analogy with twisted loop Lie
algebras. Let $a$ be an automorphism of a Lie group $G$ and $E$ be the
quotient space of the direct product $\bbR \times G$ by the equivalence
relation which identifies $(\sigma, g)$ with $(\sigma + 2 \pi, a(g))$.
Defining the projection $\pi: E \to S^1$ by the relation 
\[
\pi([(\sigma, g)]) = \rme^{\rmi \sigma},
\]
we obtain a smooth fiber bundle $E \stackrel{\pi\;}{\to} S^1$ with
fiber~$G$. Endow the space of smooth sections of this bundle with the
structure of a group defining the group composition pointwise. This group
is called the {\em twisted loop group\/} of $G$ and denoted $\calL_a(G)$.
Similarly, as for the case of $\calL(\gothg)$, one endows $\calL_a(G)$ 
with the structure of an infinite dimensional manifold modelled on the
Fr\'echet space $\calL_a(\gothg)$, where we denote the automorphism
of $\gothg$ induced by the automorphism of $G$ by the same letter $a$. One
can verify that in such a way $\calL_a(G)$ becomes a Lie group with the Lie
algebra $\calL_a(\gothg)$.

Recall that for any $g \in G$ the mapping $\Int(g): h \in G \mapsto g h
g^{-1} \in G$ is an automorphism of $G$. Such automorphisms are called {\em
inner\/} and form a normal subgroup of the group $\Aut G$. Similarly, as
for the twisted loop Lie algebras, if the automorphisms $a$ and $b$ of $G$
differ by an inner automorphism of $G$, then the twisted loop Lie groups
$\calL_a(G)$ and $\calL_b(G)$ are naturally isomorphic. Therefore, for the 
case of a semisimple Lie group $G$ one can consider only twisted loop
groups $\calL_a(G)$ where $a^K = \id_G$ for some positive integer $K$. 

One can show that there is a bijective correspondence between elements of
$\calL_a(G)$ and twisted periodic mappings from $\bbR$ to $G$. We denote by
$\tilde \gamma$ the twisted periodic mapping from $\bbR$ to $G$
corresponding to the element $\gamma \in \calL_a(G)$. Let $G$ be a
semisimple Lie group, and $a$ be an automorphism of $G$ such that $a^K =
\id_G$ for some positive integer $K$. The transformation $\sigma \to K
\sigma$ induces an injective homomorphism from $\calL_a(G)$ to $\calL(G)$
whose image is formed by the elements $\gamma$ satisfying the condition
\[
\gamma(\varepsilon_K p) = a (\gamma(p)),
\]
and will be denoted by $\calL_{a, K}(G)$. For the corresponding mapping
$\tilde \gamma$ from $\bbR$ to $G$ the above condition becomes
\[
\tilde \gamma(\sigma + 2 \pi/K) = a (\tilde \gamma(\sigma)).
\]
Thus, when $G$ is a semisimple Lie group the twisted loop group
$\calL_a(G)$ can be identified with a subgroup of $\calL(G)$.

\section{Automorphisms of twisted loop Lie algebras}

In this section $\gothg$ is always a complex simple Lie algebra and $a$ is
an automorphism of $\gothg$. As was shown in Section \ref{s:2}, studying
the twisted loop Lie algebra $\calL_a(\gothg)$, one can assume without any
loss of generality that $a^K = \id_\gothg$ for some positive integer $K$
and consider instead of $\calL_a(\gothg)$ the corresponding subalgebra
$\calL_{a,K}(\gothg)$ of $\calL(\gothg)$.

A linear homeomorphism $A$ from a Fr\'echet Lie algebra $\gothG$ to itself
is said to be an {\em automorphism\/} of $\gothG$ if
\[
A \, [\, \xi, \eta \,] = [\, A \, \xi, A \, \eta \,]
\]
for any $\xi, \eta \in \gothG$. Treating $\gothG$ as a smooth manifold, we
see that since $A$ is linear and continuous, it is smooth.

There are two main classes of  automorphisms of the Fr\'echet Lie algebra
$\calL_{a, K}(\gothg)$. The automorphisms of the first class are generated
by diffeomorphisms of $S^1$.

Let us recall that the group $\Diff(S^1)$ of smooth diffeomorphisms of the
circle $S^1$ can be supplied with the structure of a smooth infinite
dimensional manifold in such a way that it becomes a Lie group. Necessary 
information on groups of diffeomorphisms of compact manifolds and some 
relevant references are given in Appendix \ref{a:dif}. The Lie algebra
of the Lie group $\Diff(S^1)$ is the Lie algebra $\Der C^\infty(S^1)$ of
smooth vector fields on $S^1$. Here the one-parameter subgroup associated
with a vector field $X$ is actually the flow generated by $X$.

Let $f$ be a diffeomorphism of $S^1$. Consider a linear continuous mapping
$A_f: \calL_{a, K}(\gothg) \to \calL(\gothg)$ defined by the equality
\[
A_f \xi = \xi \circ f^{-1}.
\]
It is easy to see that if $\eta = A_f \xi$, then
\[
\eta(f(\varepsilon_K f^{-1}(p))) = a(\eta(p)).
\]
Hence, if
\[
f(\varepsilon_K p) = \varepsilon_K f(p)
\]
for any $p \in S^1$, then $A_f$ can be considered as a mapping from
$\calL_{a, K}(\gothg)$ to $\calL_{a, K}(\gothg)$. In this case $A_f$ is an
automorphism of $\calL_{a, K}(\gothg)$. Conversely, if the mapping $A_f$ is
a mapping from $\calL_{a, K}(\gothg)$ to $\calL_{a, K}(\gothg)$, then $f$
satisfies the above condition and $A_f$ is an automorphism of $\calL_{a,
K}(\gothg)$.

One can show that the diffeomorphisms satisfying the condition
$f(\varepsilon_K p) = \varepsilon_K f(p)$ form a Lie subgroup of the Lie
group $\Diff(S^1)$. We denote it by $\Diff_K(S^1)$. The Lie algebra of
$\Diff_K(S^1)$ is the subalgebra of $\Der C^\infty(S^1)$ formed by the
vector fields $X$ such that
\[
(X(\varphi))(\varepsilon_K p) = (X(\varphi))(p)
\]
for any function $\varphi \in C^\infty(S^1)$ satisfying the condition 
\[
\varphi(\varepsilon_K p) = \varphi(p).
\]
Denote this subalgebra by $\Der_K C^\infty(S^1)$. It is clear that we have
a left action of $\Diff_K(S^1)$ on $\calL_{a, K}(\gothg)$ realised by
automorphisms of $\calL_{a, K}(\gothg)$. Since this action is effective, we
can say that the group of automorphisms $\Aut \calL_{a, K}(\gothg)$ has a
subgroup which can be identified with the Lie group $\Diff_K(S^1)$.

The Lie group $\Diff_K(S^1)$ can be identified with a subgroup of the group
$\Diff(\bbR)$. To construct this identification we start with consideration
of general smooth mappings from $S^1$ to $S^1$.

For any $f \in C^\infty(S^1, S^1)$ one can find a smooth mapping 
$\tilde f \in C^\infty(\bbR, \bbR)$, connected with $f$ by the 
equality 
\[
f(\rme^{\rmi \sigma}) = \rme^{\rmi \tilde f(\sigma)}.
\]
The function $\tilde f$ satisfies the relation
\[
\tilde f(\sigma + 2\pi) - \tilde f(\sigma) = 2 \pi k,
\]
where $k$ is an integer, called the {\em degree\/} of $f$. From the other
hand, any smooth mapping $\tilde f \in C^\infty(\bbR, \bbR)$ which
satisfies the above relation induces a smooth mapping from $S^1$ to $S^1$.
It is evident that two functions differing by a multiple of $2 \pi$ induce
the same mapping.

If $f$ is a diffeomorphism, then its degree is $1$ for an orientation
preserving mapping, and it is $-1$ for an orientation reversing mapping.
Note that in this case the corresponding function $\tilde f$ is
strictly monotonic, and that any smooth strictly monotonic function
satisfying the relation
\[
\tilde f(\sigma + 2\pi) - \tilde f(\sigma) = \pm 2 \pi,
\]
induces a diffeomorphism of $S^1$.

If $f \in \Diff_K(S^1)$ one obtains that 
\[
\tilde f(\sigma + 2 \pi/K) = \tilde f(\sigma) + 2 \pi/K,
\]
for $K \ge 2$ and that
\[
\tilde f(\sigma + \pi) = \tilde f(\sigma) \pm \pi
\]
for $K=2$. Note that if $\xi$ is an element of $\calL_{a, K}(\gothg)$ and
$f \in \Diff_K(S^1)$ then
\[
\widetilde {A_f \xi} = \tilde \xi \circ \tilde f^{-1},
\]
where $A_f$ is the automorphism of $\calL_{a, K}(\gothg)$ induced by $f$.

The second interesting class of automorphisms of $\calL_{a, K}(\gothg)$
is formed by automorphisms generated by automorphisms of $\gothg$ acting on
the elements of $\calL_{a, K}(\gothg)$ pointwise.

Let $\alpha$ be an element of the Lie group $\calL(\Aut \gothg)$.
Consider a linear mapping from $\calL_{a, K}(\gothg)$ to $\calL(\gothg)$
defined by the equality
\[
A_\alpha \xi = \alpha \, \xi,
\]
where
\[
(\alpha \, \xi)(p) = \alpha(p)(\xi(p)).
\]
It is clear that $A_\alpha$ is a homomorphism from $\calL_{a, K}(\gothg)$
to $\calL(\gothg)$. Moreover, if $\alpha$ satisfies the relation
\[
\alpha(\varepsilon_K p) = a \,\alpha(p) a^{-1},
\]
then the mapping $A_\alpha$ is an automorphism of $\calL_{a, K}(\gothg)$.
In other words, any element of the Lie group $\calL_{\Int(a), K}(\Aut
\gothg)$
induces an automorphism of the Lie algebra $\calL_{a, K}(\gothg)$, and we
have a left action of $\calL_{\Int(a), K}(\Aut \gothg)$ on $\calL_{a,
K}(\gothg)$ realised by automorphisms of $\calL_{a, K}(\gothg)$. This
action is again effective and, therefore, $\Aut \calL_{a, K}(\gothg)$ has a
subgroup which can be identified with the Lie group $\calL_{\Int(a),
K}(\Aut \gothg)$.

Actually, if for $f \in \Diff_K(S^1)$ and $\alpha \in \calL_{\Int(a),
K}(\Aut \gothg)$ we define the automorphism $A_{(f, \alpha)}$ of $\calL_{a,
K}(\gothg)$ by
\[
A_{(f, \alpha)} \xi = \alpha(\xi \circ f^{-1}),
\]
we obtain a left effective action of the semidirect product $\Diff_K(S^1) 
\ltimes \calL_{\Int(a), K}(\Aut \gothg)$ on $\calL_{a, K}(\gothg)$ realised
by automorphisms of $\calL_{a, K}(\gothg)$. Here the group operations in
$\Diff_K(S^1) \ltimes \calL_{\Int(a), K}(\Aut \gothg)$ are given by
\[
(f_1, \alpha_1)(f_2, \alpha_2) = (f, \alpha),
\]
where
\[
f = f_1 \circ f_2, \qquad \alpha = \alpha_1^{} (\alpha_2^{} \circ
f_1^{-1}),
\]
and
\[
(f, \alpha)^{-1} = (f^{-1}, \alpha^{-1} \circ f^{-1}).
\]
Thus, we see that $\Diff_K(S^1) \ltimes \calL_{\Int(a), K}(\Aut \gothg)$
can be identified with a subgroup of the group $\Aut \calL_{a, K}(\gothg)$.
In fact, this subgroup exhausts the whole group $\Aut \calL_{a,
K}(\gothg)$.

\begin{theorem} \label{t:1}
The group of automorphisms of $\calL_{a, K}(\gothg)$ can be naturally
identified with the semidirect product $\Diff_K(S^1) \ltimes
\calL_{\Int(a), K}(\Aut \gothg)$.
\end{theorem}

\begin{proof} 
The main idea of the proof is borrowed from \cite{PreSea86}. Let $A$ be an
automorphism of $\calL_{a, K}(\gothg)$. Fix a point $p \in S^1$ and
consider the mapping $A_p$ from $\calL_{a, K}(\gothg)$ to $\gothg$ defined
by the equality
\[
A_p (\xi) = (A \xi)(p).
\]
This mapping is linear and continuous. Some necessary information on such
mappings are given in Appendix \ref{a:dis}. Certainly, $A_p$ is a
homomorphism from $\calL_{a, K}(\gothg)$ to $\gothg$.

Let $m$ be a nonnegative integer. Denote by $\chi_p^m$ a smooth function on
$S^1$ such that
\[
\chi_p^{m(m)}(p) = 1, \qquad \chi_p^{m(k)}(p) = 0, \quad k \ne m,
\]
and $\supp \chi_p^m \cap \varepsilon_K \supp \chi_p^m = \varnothing$. Let
$x$ be an arbitrary element of $\gothg$. It is not difficult to get
convinced that for any nonnegative integer $m$ the mapping
\[
\eta_{p,x}^m = \sum_{l=0}^{K-1} \chi_{\varepsilon_K^l p}^m \, a^{-l}(x)
\]
is an element of $\calL_{a, K}(\gothg)$ satisfying the conditions
\[
\eta_{p,x}^{m(m)}(p) = x, \qquad \eta_{p,x}^{m(k)}(p) = 0, \quad k \ne m.
\]

The linear mapping $A$ is invertible by definition. Therefore, there is an
element $\xi_{p,x}^0 \in \calL_{a, K}(\gothg)$ such that $A(\xi_{p,x}^0) =
\eta_{p,x}^0$. This implies that $A_p(\xi_{p,x}^0) = x$. Thus the mapping
$A_p$ is surjective.

For any open set $U \subset S^1$ the set
\[
\calL^U_{a,K}(\gothg) = \{\xi \in \calL_{a, K}(\gothg) \mid \supp \xi
\subset U\}
\]
is an ideal of $\calL_{a, K}(\gothg)$. If an element $x \in \gothg$ belongs
to the image of the restriction of $A_p$ to $\calL^U_{a,K}(\gothg)$, then
there is an element $\xi \in \calL^U_{a,K}(\gothg)$ such that $A_p(\xi) =
x$. Since $A_p$ is surjective, it follows that for any element $y \in
\gothg$ one can find an element $\eta \in \calL_{a, K}(\gothg)$ such that
$A_p(\eta) = y$. Since $[\xi, \eta]$ belongs to $\calL^U_{a,K}(\gothg)$, it
follows that $[x, y] = A_p([\xi, \eta])$ belongs to the image of
$A_p|_{\calL^U_{a,K}(\gothg)}$. This means that the image of
$A_p|_{\calL^U_{a,K}(\gothg)}$ is an ideal of $\gothg$. As the Lie algebra
$\gothg$ is simple, the mapping $A_p|_{\calL^U_{a,K}(\gothg)}$ is either
trivial or surjective.

The support of the mapping $A_p$ is the union of sets of the form
$\mbar{\{q\}\!}\,$, where $q \in S^1$ (see Appendix \ref{a:dis}). Suppose
that $\supp A_p = \mbar{\{q\}\!} \cup \mbar{\{q'\}\!}$ and $\mbar{\{q\}\!}
\cap \mbar{\{q'\}\!} = \varnothing$. Let $U$ and $U'$ be disjoint
neigbourhoods of $\mbar{\{q\}\!}$ and $\mbar{\{q'\}\!}$ respectively. Since
$S^1$ is a normal topological space such neighbourhoods do exist.
It is clear that $\calL^U_{a, K}(\gothg)$ and $\calL^{U'}_{a, K}(\gothg)$
are commuting ideals of $\calL_{a, K}(\gothg)$. Therefore, the images
$A_p|_{\calL^U_{a, K}(\gothg)}$ and $A_p|_{\calL^{U'}_{a, K}(\gothg)}$ are
commuting ideals of $\gothg$. Hence, one of the mappings $A_p|_{\calL^U_{a,
K}(\gothg)}$ and $A_p|_{\calL^{U'}_{a, K}(\gothg)}$ is surjective, the
other one is trivial. Thus, the support of $A_p$ has the form
$\mbar{\{f'(p)\}\!}\,$ for some mapping $f': S^1 \to S^1$, and we can write 
\[
A_p(\xi) = \sum_{m=0}^M c_p^m (\xi^{(m)}(f'(p))),
\]
for some nonnegative integer $M$ and endomorphisms $c_p^m$ (see Appendix
\ref{a:dis}). We assume that the endomorphisms $c_p^m$ are defined for all
nonnegative $m$, but $c_p^m = 0$ for $m > M$.

It is clear that
\[
A_p^{}(\eta_{f'(p),x}^m) = c_p^m(x).
\]
Using the relations
\[
[\eta_{p,x}^m, \eta_{p,y}^n]^{(m+n)}(p) = {m + n \choose m} [x, y], \qquad
[\eta_{p,x}^m, \eta_{p,y}^n]^{(k)}(p) = 0, \quad k \ne m+n,
\]
we obtain
\[
A_p^{}([\eta_{f'(p),x}^m, \eta_{f'(p),y}^n]) = {m + n \choose m} c_p^{m+n}
([x, y]).
\]
Since $A_p$ is a homomorphism, we have
\[
A_p^{}([\eta_{f'(p),x}^m, \eta_{f'(p),y}^n]) = [A_p(\eta_{f'(p),x}^m),
A_p(\eta_{f'(p),y}^n)],
\]
therefore,
\[
A_p([\eta_{f'(p),x}^m, \eta_{f'(p),y}^n]) = [c_p^m(x), c_p^n(y)].
\]
Thus, one has the equalities
$$
{m + n \choose m} c_p^{m+n} ([x, y]) = [c_p^m(x), c_p^n(y)]. \eqno{(*)}
$$
In particular, for $m = n = 0$ the equality
$$
c_p^0([x, y]) = [c_p^0(x), c_p^0(y)] \eqno{(**)}
$$
is valid. Since $\gothg$ is simple, the mapping $c_p^0$ is either trivial
or surjective. Suppose that it is trivial. Putting in the equality (*) 
$n = 0$, we obtain
\[
c_p^m([x, y]) = [c_p^m(x), c_p^0(y)].
\]
Since the Lie algebra $\gothg$ is simple, then $[\gothg, \gothg] = \gothg$.
Therefore, for any $m$ the mapping $c_p^m$ is trivial. Hence, the mapping
$A_p$ is also trivial. This contradicts surjectivity of $A_p$. Thus,
$c_p^0$ is surrjective, and the equality $(**)$ says that it is an
automorphism of the Lie algebra $\gothg$.

Putting in $(*)$ $m = 0$ and $n = 1$, we obtain
\[
c_p^1([x, y]) = [c_p^0(x), c_p^1(y)]
\]
for any $x, y \in \gothg$. Rewrite this equality as
\[
(c_p^0)^{-1} (c_p^1([x, y])) = [x, (c_p^0)^{-1}(c_p^1(y))].
\]
Therefore,
\[
((c_p^0)^{-1} c_p^1) \ad (x) = \ad (x) ((c_p^0)^{-1} c_p^1)
\]
for any $x \in \gothg$. Since the Lie algebra $\gothg$ is simple, the
linear operator $(c_p^0)^{-1} c_p^1$ is multiplication by some scalar,
denote it by $\rho$. Thus, we have $c_p^1 = \rho \, c_p^0$. Relation $(*)$
for $m = 1$ and $n = 1$ takes the form
\[
2 \, c_p^2([x, y]) = [c_p^1(x), c_p^1(y)] = \rho^2 c_p^0([x, y]).
\]
Therefore, $c_p^2 = (\rho^2/2) c_p^0$. In general case we have $c_p^m
= (\rho^m/m!) c_p^0$ for any positive $m$. From the other hand, $c_p^m = 0$
for $m > M$. It is possible only if $\rho = 0$. Hence, $c_p^m = 0$ for all
$m > 0$.

Define a mapping $\alpha: S^1 \to \Aut \gothg$ by
\[
\alpha(p) = c_p^0,
\]
then one can write
\[
A \, \xi = \alpha (\xi \circ f').
\]
Since for any $\xi \in \calL_{a, K}(\gothg)$ the mapping $A \, \xi$ belongs
to $\calL_{a, K}(\gothg)$, the mappings $f'$ and $\alpha$ must be smooth.
The mapping $f'$ is actually an element of $\Diff_K(S^1)$, and $\alpha$
belongs to $\calL_{\Int(a), K}(\Aut \gothg)$. Hence, defining $f =
f^{\prime -1}$, we see that
\[
A \, \xi = \alpha (\xi \circ f^{-1}).
\]
Thus, an arbitrary automorphism of $\calL_{a, K}(\gothg)$ has the above
form for some $f \in \Diff_K(S^1)$ and some~$\alpha \in \calL_{\Int(a),
K}(\Aut \gothg)$.
\end{proof}

Note that in the case where $\gothg$ is a complex Lie algebra, the Lie
group $\Aut \gothg$ is a complex Lie group. In this case
$\calL_{\Int(a),K}(\Aut \gothg)$ is also a complex Lie group. From the
other hand, any diffeomorphism from the identity component of the Lie group
$\Diff_K(S^1)$ to a complex Lie group is trivial (see, for example,
\cite{PreSea86}). This implies that $\Diff_K(S^1)$ cannot be endowed with
the structure of a complex Lie group. Therefore, even in the case where
$\gothg$ is a complex Lie algebra we consider $\calL_{\Int(a),K}(\Aut
\gothg)$ as a real Lie group. Thus, the identification described in Theorem
\ref{t:1} supplies the group $\Aut \calL_{a, K}(\gothg)$ with the structure
of a real Lie group. Here the action of the group $\Aut \calL_{a,
K}(\gothg)$ on $\calL_{a, K}(\gothg)$, where $\calL_{a, K}(\gothg)$ is
treated as a real manifold, is
smooth.

The Lie algebra of the Lie group $\Aut \gothg$ is the Lie algebra $\Der
\gothg$ of derivations of $\gothg$. The situation is almost the same for
the case of the Lie group $\Aut \calL_{a,K}(\gothg)$. Actually, any element
of the Lie algebra of the Lie group $\Aut \calL_{a,K}(\gothg)$ induces a
derivation of $\calL_{a,K}(\gothg)$, but in the case where $\gothg$ is a
complex Lie algebra there are derivations of $\calL_{a,K}(\gothg)$ which
cannot be obtained in such a way. To show this, let us consider first the
Lie algebra of $\Aut \calL_{a,K}(\gothg)$. Using the identification
described in Theorem \ref{t:1}, we see that this Lie algebra can be
identified with the semidirect product of the Lie algebra of the Lie group
$\Diff_K(S^1)$ and the Lie algebra of the Lie group $\calL_{\Int(a),
K}(\Aut \gothg)$. As we already noted the Lie algebra of $\Diff_K(S^1)$ is
the subalgebra $\Der_K C^\infty(S^1)$ of the Lie algebra $\Der
C^\infty(S^1)$ of smooth vector fields on $S^1$. The Lie algebra of the Lie
group $\calL_{\Int(a), K}(\Aut \gothg)$ is $\calL_{\Ad(a), K}(\Der
\gothg)$. Thus, the Lie algebra of the group of automorphisms of $\calL_{a,
K}(\gothg)$ can be naturally identified with the Lie algebra $\Der_K
C^\infty (S^1) \ltimes \calL_{\Ad(a), K}(\Der \gothg)$.

By a {\em derivation\/} of a Fr\'echet Lie algebra $\gothG$ we mean a
continuous linear mapping $D$ from $\gothG$ to $\gothG$ which satisfies the
relation
\[
D [\xi, \eta] = [D \xi, \eta] + [\xi, D \eta].
\]
Note again that continuity and linearity imply smoothness.

The derivation of $\calL_{a,K}(\gothg)$ corresponding to an element of
$\Der_K C^\infty (S^1) \ltimes \calL_{\Ad(a), K}(\Der \gothg)$ is
constructed as follows. Define the action of a vector field 
$X \in \Der_K (S^1)$ on an element  $\xi \in \calL_{a, K}(\gothg)$ 
in the usual way. Let $(e_i)$ be a basis of $\gothg$, then for any 
element $\xi \in \calL_{a, K}(\gothg)$ one can write 
\[
\xi = \sum_i e_i \xi^i,
\]
where $\xi^i$ are smooth functions on $S^1$. Then one assumes that
\[
X(\xi) = \sum_i e_i X(\xi^i).
\]
One can get convinced that this definition does not depend on the choice of
a basis $(e_i)$. Let $(X, \delta)$ be an element of $\Der_K C^\infty (S^1)
\ltimes \calL_{\Ad(a), K}(\Der \gothg)$. Consider the corresponding
one-parameter subgroup of the Lie group $\Diff_K(S^1) \ltimes
\calL_{\Int(a), K}(\Aut \gothg)$. It is determined by two mappings
$\lambda: \bbR \to \Diff_K(S^1)$ and $\theta: \bbR \to \calL_{\Int(a),
K}(\Aut \gothg)$. For any fixed element $\xi \in \calL_{a, K}(\gothg)$ one
has a curve $\tau \in \bbR \mapsto \theta(\tau) (\xi \circ
(\lambda(\tau))^{-1})$ in $\calL_{a, K}(\gothg)$. The tangent vector to
this curve at zero can be treated as the action of a linear operator $D$ on
the element $\xi$. It is clear that
\[
D \xi = - X(\xi) + \delta (\xi),
\]
where
\[
(\delta(\xi))(p) = \delta(p) (\xi(p)).
\]
One can verify that $D$ is a derivation of the Lie algebra~$\calL_{a,
K}(\gothg)$. In can be shown also that in the case where $\gothg$ is a real
Lie algebra the derivations of the above form exhaust all possible
derivations of the Lie algebra $\calL_{a, K}(\gothg)$. In the case where
$\gothg$ is a complex Lie algebra to exhaust all derivations one should
assume that the vector field $X$ may be complex.

\section{$\bbZ$-gradations of twisted loop Lie algebras}

In general, dealing with $\bbZ$-gradations of infinite dimensional Lie
algebras we confront with necessity to work with infinite series of their
elements, or, in other words, with series in Fr\'echet spaces. The relevant
information on such series is given in Appendix \ref{a:ser}.

Let $\gothG$ be a Fr\'echet Lie algebra. Suppose that for any $k \in \bbZ$
there is given a closed subspace $\gothG_k$ of $\gothG$ such that

(a) for any $k, l \in \bbZ$ one has $[\gothG_k, \gothG_l] \subset
\gothG_{k+l}$,

(b) any element $\xi$ of $\gothG$ can be uniquely represented as
an absolutely convergent series
\[
\xi = \sum_{k \in \bbZ} \xi_k,
\]
where $\xi_k \in \gothG_k$. In this case we say that the Fr\'echet Lie
algebra $\gothG$ is supplied with a {\em $\bbZ$-gradation\/}, and
call the subspaces $\gothG_k$ the {\em grading subspaces\/} of
$\gothG$ and the elements $\xi_k$ the {\em grading components\/} of $\xi$.
If $F$ is an isomorphism from the Fr\'echet Lie algebra $\gothG$ to a
Fr\'echet Lie algebra $\gothH$, then taking the subspaces $\gothH_k =
F(\gothG_k)$ of $\gothH$ as grading subspaces we endow $\gothH$ with a
$\bbZ$-gradation. In this case we say that the $\bbZ$-gradations of
$\gothG$ and $\gothH$ under consideration are {\em conjugated\/} by the
isomorphism $F$. It is clear that if the grading components of an element
$\xi \in \gothG$ are $\xi_k$, then the grading components $F(\xi)_k$ of the
element $F(\xi) \in \gothH$ are $F(\xi_k)$.

As the simplest example, let us consider the so-called {\em standard
gradation\/} of $\calL(\gothg)$. Denote by $\lambda$ the standard
coordinate function on $\bbC$ and its restriction to $S^1$. The grading
subspaces for the standard gradation are defined as
\[
\calL(\gothg)_k = \{\lambda^k x \mid x \in \gothg\},
\]
and the expansion of a general element $\xi$ of $\calL(\gothg)$ over
grading subspaces is the representation of $\xi$ as a Fourier
series:
\[
\xi = \sum_{k \in \bbZ} \lambda^k x_k,
\]
that in terms of the mapping $\tilde \xi$ has the usual form
\[
\tilde \xi = \sum_{k \in \bbZ} \rme^{i k s} x_k,
\]
with
\[
x_k = \frac{1}{2 \pi} \int_{[0,2\pi]} \rme^{-\rmi k s} \, \tilde \xi
\, \rmd s.
\]
{}From the theory of Fourier series it follows that the Fourier series of
any
element $\xi \in \calL(\gothg)$ converges absolutely to $\xi$ as a series
in the Fr\'echet space $\calL(\gothg)$. Hence, we really have a
$\bbZ$-gra\-dation of $\calL(\gothg)$. 

The necessity to include the requirement of absolute convergence in the
definition of $\bbZ$-gra\-dation is justified by the following proposition.

\begin{proposition} \label{p:2}
Let a Fr\'echet Lie algebra $\gothG$ be supplied with a $\bbZ$-gradation.
For any two elements of $\gothG$,
\[
\xi = \sum_{k \in \bbZ} \xi_k, \qquad \eta = \sum_{k \in \bbZ} \eta_k,
\]
the grading components of $[\xi, \eta]$ are given by
\[
[\xi, \eta]_k = \sum_{l \in \bbZ} [\xi_{k-l}, \eta_l].
\]
Here the series at the right hand side converges absolutely.      
\end{proposition}

\begin{proof}
First prove that the series $\sum_{(k,l) \in \bbZ \times \bbZ} [\xi_k,
\eta_l]$ converges absolutely. Let $\alpha$ be an element of $\calD(\bbZ
\times \bbZ)$, fix a positive integer $m$, and define
\[
r_{\alpha,m} = \sum_{(k,l) \in \alpha} \| [\xi_k, \eta_l] \|_m.
\]
There are elements $\beta, \gamma \in \calD(\bbZ)$ such that $\alpha
\subset \beta \times \gamma$. Using Lemma \ref{l:1}, we obtain
\begin{eqnarray*}
&& r_{\alpha,m} \leq \sum_{(k,l) \in \beta \times \gamma}  \| [\xi_k,
\eta_l] \|_m \leq C_m \sum_{(k,l) \in \beta \times \gamma} \| \xi_k
\|_m \|
\eta_l \|_m \\
&& \hspace{6em} {} = C_m \left( \sum_{k \in \beta} \| \xi_k \|_m
\right)
\left( \sum_{l \in \gamma} \| \eta_l \|_m \right) \leq C_m \left(
\sum_{k
\in \bbZ} \| \xi_k \|_m \right) \left( \sum_{l \in \bbZ} \| \eta_l
\|_m
\right).
\end{eqnarray*}
It is clear that for any positive integer $m$ the net
$(r_{\alpha,m})_{\alpha \in \calD(\bbZ)}$ is monotonically increasing,
that means that $r_{\alpha,m} \ge r_{\beta,m}$ if $\alpha \succcurlyeq
\beta$. The above inequalities show that it is also bounded above.
Similarly as it is for the case of sequences, such a net is convergent.
Therefore, the series $\sum_{(k,l) \in \bbZ \times \bbZ} [\xi_k, \eta_l]$
converges absolutely.

As follows from Proposition \ref{p:1} one can write
\[
\sum_{(k,l) \in \bbZ \times \bbZ} [\xi_k, \eta_l] = \sum_{k \in \bbZ}
\left( \sum_{l \in \bbZ}[\xi_k, \eta_l] \right).
\]
For a fixed $k$ the net $\left( \sum_{l \in \alpha} [\xi_k, \eta_l]
\right)_{\alpha \in \calD(\bbZ)}$ converges absolutely. It is clear that
this net coincides with the net $\left( [\xi_k, \sum_{l \in \alpha} \eta_l]
\right)_{\alpha \in \calD(\bbZ)}$. Since the net $\left( \sum_{l \in
\alpha} \eta_l \right)_{\alpha \in \calD(\bbZ)}$ converges to $\eta$ and
the Lie algebra operation in $\gothG$ is continuous, one has
\[
\sum_{l \in \bbZ} [\xi_k, \eta_l] = [\xi_k, \eta].
\]
Similarly, one obtains
\[
\sum_{k \in \bbZ} [\xi_k, \eta] = [\xi, \eta].
\]
Using again Proposition \ref{p:1}, we come to the equality
\[
[\xi, \eta] = \sum_{k \in \bbZ} \sum_{l \in \bbZ} [\xi_{k-l}, \eta_l],
\]
where for any $k$ the series $\sum_{l \in \bbZ} [\xi_{k-l}, \eta_l]$
converges absolutely.
\end{proof}

Suppose that a Fr\'echet Lie algebra $\gothG$ is supplied with
a $\bbZ$-gradation such that for any element $\xi = \sum_{k \in \bbZ}
\xi_k$ of $\gothG$ the series $\sum_{k \in \bbZ} k \, \xi_k$ converges
unconditionally. In this case one can define a linear operator $Q$ in
$\gothG$, acting on an element  $\xi = \sum_{k \in \bbZ} \xi_k$ as
\[
Q \xi = \sum_{k \in \bbZ} k \, \xi_k.
\]
Actually the elements $k \,\xi_k$ are the grading components of the element
$Q \xi$, therefore, the series $\sum_{k \in \bbZ} k \, \xi_k$ converges
absolutely by the definition of a $\bbZ$-gradation. It is clear that
\[
\gothG_k = \{ \xi \in \gothG \mid Q \xi = k \xi \}.
\]
We call the linear operator $Q$ the {\em grading operator\/} and say that
the $\bbZ$-gradation under consideration is {\em generated by grading
operator\/}. If a $\bbZ$-gradation of a Fr\'echet Lie algebra $\gothG$ and
a $\bbZ$-gradation of a Fr\'echet Lie algebra $\gothH$ are conjugated by an
isomorphism $F$, and the $\bbZ$-gradation of $\gothG$ is generated by a 
grading operator $Q$, then the $\bbZ$-gradation of $\gothH$ is generated by
the grading operator $F Q F^{-1}$.

The standard gradation of $\calL(\gothg)$ is generated by a grading
operator $Q$ such that
\[
\widetilde{Q \xi} = - \rmi \, \rmd \tilde \xi / \rmd s.
\]
Here the operator $Q$ is a derivation of $\calL(\gothg)$. In general we
have the following statement.

\begin{proposition} 
Let a $\bbZ$-gradation of a Fr\'echet Lie algebra $\gothG$ be generated by
a grading operator $Q$. The equality
\[
Q \, [\xi, \eta] = [Q \, \xi, \eta] + [\xi, Q \, \eta].
\]
is valid for any $\xi, \eta \in \gothG$.
\end{proposition}

\begin{proof}
Using Proposition \ref{p:2}, one obtains
\[
Q \, [\xi, \eta] = \sum_{k \in \bbZ} (Q \, [\xi, \eta])_k = \sum_{k
\in \bbZ} k [\xi, \eta]_k = \sum_{k \in \bbZ} \left( \sum_{l \in \bbZ}
k [\xi_{k-l}, \eta_l] \right).
\]
In a similar way one comes to the equalities
\[
[Q \, \xi, \eta] = \sum_{k \in \bbZ} \left( \sum_{l \in \bbZ} (k-l)
[\xi_{k-l}, \eta_l] \right), \qquad  [\xi, Q \, \eta] = \sum_{k \in \bbZ}
\left( \sum_{l \in \bbZ} l [\xi_{k-l}, \eta_l] \right).
\]
The three above equalities imply the validity of the statement of the
proposition.
\end{proof}
It follows from this lemma that if the grading operator $Q$ generating a
$\bbZ$-gradation of a Fr\'echet Lie algebra $\gothG$ is continuos, it is
a derivation of $\gothG$.

We call a $\bbZ$-gradation of a Fr\'echet Lie algebra $\gothG$
{\em integrable\/} if the mapping $\slantPhi$ from $\bbR \times
\gothG$ to $\gothG$ defined by the relation
\[
\slantPhi(\tau, \xi) = \sum_{k \in \bbZ} \rme^{- \rmi k \tau} \xi_k
\]
is smooth. Here as usually we denote by $\xi_k$ the grading components of
the element $\xi$ with respect to the $\bbZ$-gradation under consideration.

For each fixed $\xi \in \gothG$ the mapping $\slantPhi$ induces a smooth 
curve $\slantPhi_\xi: \bbR \to \gothG$ given by the equality
\[
\slantPhi_\xi(\tau) = \slantPhi(\tau, \xi).
\]

\begin{proposition}
Any integrable $\bbZ$-gradation of a Fr\'echet Lie algebra $\gothG$ is
generated by grading operator. The corresponding grading operator $Q$ acts
on an element $\xi \in \gothG$ as
\[
Q \, \xi = \left. \rmi \frac{\rmd}{\rmd \, t} \right|_0 \slantPhi_\xi,
\]
where we denote by $t$ the standard coordinate function on $\bbR$.
\end{proposition}

\begin{proof}
Since the mapping $\slantPhi$ is smooth and linear in $\xi$, then $Q$ is 
a continuous linear operator on $\gothG$. Therefore, for any net
$(\xi_\alpha)_{\alpha \in \calD(\bbZ)}$ in $\gothG$ which converges to an
element $\xi \in \gothG$, the net $(Q \,\xi_\alpha)_{\alpha \in
\calD(\bbZ)}$ converges to $Q \, \xi$. The net $(\xi_\alpha)_{\alpha \in
\calD(\bbZ)}$, where
\[
\xi_\alpha = \sum_{k \in \alpha} \xi_k,
\]
where $\xi_k$ are the grading components of $\xi$, converges to $\xi$.
Since for any $\alpha \in \calD(\bbZ)$ the element $\xi_\alpha$ is the sum
of a finite number of grading components, one has
\[
Q \, \xi_\alpha = \left. \rmi \frac{\rmd}{\rmd \, t} \right|_0
\slantPhi_{\xi_\alpha} = \sum_{k \in \alpha} k \, \xi_k.
\]
This means that $Q \xi = \sum_{k \in \bbZ} k \, \xi_k$. Thus, the linear
operator $Q$ generates the $\bbZ$-gradation under consideration.
\end{proof}

\begin{proposition} \label{p:7}
Let a Fr\'echet Lie algebra $\gothG$ be supplied with an integrable
$\bbZ$-gradation. Then for any fixed $\tau \in \bbR$ the mapping
$\xi \in \gothG \mapsto \slantPhi(\tau, \xi) \in \gothG$ is an automorphism
of $\gothG$. The mapping $\slantPhi$ satisfies the relation
\[
\slantPhi(\tau_1, \slantPhi(\tau_2, \xi)) = \slantPhi(\tau_1 +
\tau_2, \xi).
\]
\end{proposition}

\begin{proof}
{}From Proposition \ref{p:2} it follows that one can write
\[
[\slantPhi(\tau, \xi), \slantPhi(\tau, \eta)] = \sum_{k \in \bbZ} \sum_{l
\in \bbZ} [(\slantPhi(\tau, \xi))_{k-l}, (\slantPhi(\tau, \eta))_l].
\]
It is clear that
\[
(\slantPhi(\tau, \xi))_{k-l} = \rme^{- \rmi (k-l) \tau} \xi_{k-l}, \qquad
(\slantPhi(\tau, \eta))_{l} = \rme^{- \rmi l \tau} \eta_l.
\]
Therefore, one has
\[
[\slantPhi(\tau, \xi), \slantPhi(\tau, \eta)] = \sum_{k \in \bbZ}
\rme^{-\rmi k \tau} \sum_{l \in \bbZ} [\xi_{k-l}, \eta_l] = \sum_{k \in
\bbZ} \rme^{-\rmi k \tau} [\xi, \eta]_k = \slantPhi(\tau, [\xi, \eta]).
\]
That proves the first statement of the proposition. The second statement of
the proposition is evident.
\end{proof}

Let us return to consideration of twisted loop Lie algebras. Suppose that
$\gothg$ is a complex simple Lie algebra, and $a$ is an automorphism of
$\gothg$ satisfying the relation $a^K = \id_\gothg$ for some positive
integer~$K$. Assume that the twisted Lie algebra $\calL_{a, K}(\gothg)$ is
endowed with an integrable $\bbZ$-gradation. Define a mapping $\varphi$
from $\bbR$ to the Lie group $\Aut \calL_{a, K}(\gothg)$ by the equality
\[
(\varphi(\tau))(\xi) = \slantPhi(\tau, \xi).
\]
It is a curve in the Lie group $\Aut \calL_{a, K}(\gothg)$. Using the
identification of $\Aut \calL_{a, K}(\gothg)$ with the Lie group
$\Diff_K(S^1) \ltimes \calL_{\Int(a), K}(\Aut \gothg)$, for any 
$\tau \in \bbR$ one can write
\[
\varphi(\tau) = (\lambda(\tau), \theta(\tau)),
\]
where $\lambda$ is a mapping from $\bbR$ to the Lie group $\Diff_K(S^1)$
and $\theta$ is a mapping from $\bbR$ to the Lie group 
$\calL_{\Int(a), K}(\Aut \gothg)$. The mapping $\lambda$ induces a mapping 
$\slantLambda$ from $\bbR \times S^1$ to $S^1$ given by
\[
\slantLambda(\tau, p) = (\lambda(\tau))(p),
\]
and the mapping $\theta$ induces a mapping $\slantTheta$ from $\bbR \times
S^1$ to $\Aut \gothg$ given by
\[
\slantTheta(\tau, p) = (\theta(\tau))(p).
\]
Using the mappings $\slantLambda$ and $\slantTheta$ one can write
\[
(\slantPhi(\tau, \xi))(p) = \slantTheta(\tau, p)
(\xi(\slantLambda^{-1}(\tau, p))),
\]
where the mapping $\slantLambda^{-1}: \bbR \times S^1 \to S^1$ is defined
by the equality
\[
\slantLambda^{-1}(\tau, \slantLambda(\tau, p)) = p.
\]
Since the mapping $\slantPhi$ is smooth, also the mappings $\slantLambda$ 
and $\slantTheta$ are smooth. Therefore, by the exponential law (see
Appendix \ref{a:dif}), the mappings $\lambda$ and $\theta$ are also smooth.
Thus, the curve $\varphi$ is a smooth  curve in the Lie group $\Aut
\calL_{a, K}(\gothg)$. Actually, as follows from Proposition \ref{p:7}, it
is a one-parameter subgroup of $\Aut \calL_{a, K}(\gothg)$. The tangent
vector to the curve $\varphi$ at zero is a derivation of $\calL_{a,
K}(\gothg)$ which coincides with the linear operator $-\rmi \, Q$.
Therefore, one has the equality
\[
Q \, \xi = - \rmi \, X(\xi) + \rmi \, \delta (\xi).
\]
Here $X \in \Der_K C^\infty(S^1)$ is the vector field being the tangent
vector at zero to the curve $\lambda$ in $\Diff_K(S^1)$, and $\delta$ is
the tangent vector at zero to the curve $\theta$ in $\calL_{\Int(a),
K}(\Aut \gothg)$. 

Note that the mapping $\slantLambda$ corresponding to the mapping $\lambda$
is a flow on $S^1$, and $X$ is the vector field which generates this flow.

\begin{proposition} \label{p:9}
Either the vector field $X$ is zero vector field, or it has no zeros.
\end{proposition}

\begin{proof}
It is clear that $\slantPhi(\tau + 2 \pi, \xi) =
\slantPhi(\tau, \xi)$ for any $\xi \in \calL(\gothg)$. It implies that
$\slantLambda(\tau + 2 \pi, p) = \slantLambda(\tau, p)$ for any $p \in
S^1$. According to the mechanical interpretation of the flow,
$\slantLambda(\tau, p)$ is the position of a particle at time $\tau$, if
its position at zero time is $p$. Here the velocity of the particle at time
$\tau$ is $X(\Lambda(\tau, p))$. If $p \in S^1$ is a zero of $X$, then a
particle placed at the point $p$ at some instant of time will forever
remain at that point. If $X(p) \neq 0$, a particle placed at the point $p$
will instantly move in the same direction, and it cannot pass any zero of
the vector field $X$. If $X$ has zeros, this contradicts the periodicity of
$\slantLambda$ in the first argument. There is no contradiction only if $X$
is zero vector field.
\end{proof}

Recall that any derivation of a simple Lie algebra is an inner derivation.
Therefore, if $\delta$ is an element of $\calL_{\Int(a), K}(\Der \gothg)$,
then there exists a unique element $\eta$ of $\calL_{a, K}(\gothg)$ such
that
\[
\delta(\xi) = [\eta, \xi].
\]
Thus, we come to the following proposition.

\begin{proposition} \label{p:8}
The grading operator $Q$ generating an integrable $\bbZ$-gradation of a
twisted loop Lie algebra $\calL_{a, K}(\gothg)$ acts on an element $\xi \in
\calL_{a, K}(\gothg)$ as
\[
Q \, \xi = - \rmi \, X(\xi) + \rmi \, [\eta, \xi].
\]
where $X \in \Der_K C^\infty(S^1)$, and $\eta$ is an element of
$\calL_{a, K}(\gothg)$.
\end{proposition}

We will not consider $\bbZ$-gradations with infinite dimensional grading
subspaces. Therefore, the vector field $X$ cannot be zero vector field.
Indeed, suppose that $X = 0$ and $Q \, \xi = \rmi \, [\eta, \xi] = k \xi$, 
then $\| [\eta, \xi] \|_1 = |k| \|\xi\|_1$. From Lemma \ref{l:1} one
obtains
\[
|k| \le C \|\eta\|_1.
\]
Hence, we have only a finite number of grading subspace, thus, at least
some of them must be infinite dimensional.

{}From now on we identify any element $\xi$ of $\calL_{a, K}(\gothg)$ with 
the corresponding mapping $\tilde \xi$ from $\bbR$ to $\gothg$ omitting 
the tilde. Similarly, we identify each element of $\Diff_K(S^1)$ with the
corresponding mapping $\tilde f$ again omitting the tilde. An element $X$
of $\Der_K C^\infty(S^1)$ is identified with the vector field on $\bbR$,
which we denote again by $X$. One has
\[
X = v \, \rmd / \rmd s,
\]
where the function $v$ satisfies the relation
\[
v(\sigma + 2 \pi /K) = v(\sigma).
\]

\begin{proposition} \label{p:10}
Let the twisted loop Lie algebra $\calL_{a, K}(\gothg)$ be endowed with an
integrable $\bbZ$-gradation with finite dimensional grading subspaces, and
$Q$ be the corresponding grading operator, which has the form described in
Proposition \ref{p:8}. For any diffeomorphism $f \in \Diff_K S^1$ one has
\[
A_f Q A_f^{-1} \xi = - \rmi f_* X(\xi) + \rmi \, [\eta, \xi],
\]
where $A_f$ is the automorphism of $\calL_{a, K}(\gothg)$ induced by $f$.
Here the diffeomorphism $f$ can be chosen so that
\[
f_* X = \kappa \, \rmd / \rmd s
\]
for some nonzero real constant $\kappa$.
\end{proposition}

\begin{proof}
The first statement of the proposition follows from the well known equality
\[
f_* X (\varphi) = f^{-1*} X(f^* \varphi)
\]
valid for any $\varphi \in C^\infty(S^1)$.

Writing the vector field $X$ as $v \, \rmd / \rmd s$, in accordance with
Proposition \ref{p:9} we conclude that the function $v$ has no zeros.  Thus
we can consider a diffeomorphism $f$ of $\Diff_K(S^1)$ with
\[
f(\sigma) = \kappa \int_0^\sigma \frac{\rmd \sigma'}{v(\sigma')}.
\]
Here the constant $\kappa$ is fixed by $f(2 \pi/K) = 2 \pi/K$. It is
easy to verify that $f_* X = \kappa \, \rmd / \rmd s$. Thus, the second
statement of the proposition is true.
\end{proof}

Without any loss of generality one can assume that the constant $\kappa$ of
the above proposition is positive. Indeed, if it is not the case one can
do so performing the mapping $\xi(\sigma) \to \xi(-\sigma)$ which maps
$\calL_{a, K}(\gothg)$ isomorphically onto $\calL_{a^{-1}, K}(\gothg)$.

Let $G$ be a simply connected Lie group whose Lie algebra coincides with
$\gothg$. Denote the automorphism of $G$ corresponding to the automorphism
$a$ of $\gothg$ by the same letter $a$. Since $G$ is a complex
simple Lie group, we will consider it as a linear group. 
The following proposition is evident.

\begin{proposition} \label{p:11}
Let the twisted loop Lie algebra $\calL_{a, K}(\gothg)$ be endowed with an
integrable $\bbZ$-gradation, and $Q$ be the corresponding grading operator,
which has the form described in Proposition \ref{p:8}. Let $\gamma$ be a
smooth mapping from $\bbR$ to $G$ satisfying the relation
\[
\gamma(\sigma + 2\pi/K) = a(g \, \gamma(\sigma))
\]
for some $g \in G$. Consider a linear mapping $A_\gamma$ acting on any 
element $\xi \in \calL_{a, K}(\gothg)$ as
\[
A_\gamma \xi = \gamma \, \xi \gamma^{-1}.
\]
The mapping $A_\gamma$ is an isomorphism from $\calL_{a, K}(\gothg)$ to the
Lie algebra of smooth mappings $\xi$ from $\bbR$ to $\gothg$ satisfying the
equality
\[
\xi(\sigma + 2\pi/K) = a(g \, \xi(\sigma) g^{-1}).
\]
This isomorphism conjugates the $\bbZ$-gradation of $\calL_{a, K}(\gothg)$
and the $\bbZ$-gradation generated by the grading operator $A_\gamma \, Q
A_\gamma^{-1}$ which acts as
\[
A_\gamma \, Q A_\gamma^{-1} \xi = - \rmi \, X (\xi) 
+ \rmi \, [\gamma \, \eta \, \gamma^{-1} + X(\gamma) \gamma^{-1}, \xi].
\]
\end{proposition}

Now we are able to prove our main theorem.

\begin{theorem}
An integrable $\bbZ$-gradation of a twisted loop Lie algbera $\calL_{a,
K}(\gothg)$ with finite dimensional grading subspaces is conjugated by 
an isomorphism to a $\bbZ$-gradation of an appropriate twisted loop Lie
algebra $\calL_{a', K'}(\gothg)$ generated by grading operator
\[
Q' \xi = - \rmi \, \rmd \xi / \rmd s.
\]
Here the automorphisms $a$ and $a'$ differ by an inner automorphism of
$\gothg$.
\end{theorem}

\begin{proof}
In accordance with Proposition \ref{p:8} the grading operator of an
integrable $\bbZ$-gradation of $\calL_{a, K}(\gothg)$ with finite
dimensional grading subspaces is specified by the choice of a vector 
field $X \in \Der_K C^\infty(S^1)$ and by an element 
$\eta \in \calL_{a,K}(\gothg)$. Having in mind Proposition \ref{p:10} 
and the discussion given just below it, we assume without loss of 
generality that
\[
X = \kappa \, \rmd / \rmd s
\] 
for some positive real constant $\kappa$. 

Let a mapping $\gamma: \bbR \to G$ be a solution of the equation
\[
\kappa \, \gamma^{-1} \rmd \gamma / \rmd s = - \eta.
\]
It is well known that this equation always has solutions, all its solutions
are smooth, and if $\gamma$ and $\gamma'$ are two solutions then
\[
\gamma' = g \gamma
\]
for some $g \in G$. Using the equality
\[
\eta(\sigma + 2 \pi / K) = a (\eta(\sigma)),
\]
one concludes that, if $\gamma$ is a solution, then the mapping $\gamma'$
defined by the equality
\[
\gamma'(\sigma) = a^{-1} (\gamma(\sigma + 2 \pi / K))
\]
is also a solution. Hence, for some $g \in G$ one has
\[
\gamma(\sigma + 2 \pi / K) = a (g \gamma(\sigma)).
\]

The mapping $A_\gamma$, described in Proposition \ref{p:11}, accompanied 
by the transformation $\sigma \to \sigma/K$ maps $\calL_{a, K}(\gothg)$
isomorphically onto the Fr\'echet Lie algebra $\gothG$ formed by smooth
mappings $\xi$  from $\bbR$ to $\gothg$ satisfying the condition
\[
\xi(\sigma + 2 \pi) = a'(\xi(\sigma)),
\]
where $a' = a \circ \Ad(g)$. Denote the grading operator generating the
corresponding conjugated $\bbZ$-gradation again by $Q$. In accordance 
with Proposition \ref{p:11} the operator $Q$ acts on an element $\xi$ as
\[
Q \, \xi = - \rmi K' \rmd \xi / \rmd s,
\]
where $K' = \kappa K$. Suppose that for some integer $k$ the grading
subspace $\gothG_k$ is nontrivial and $\xi \in \gothG_k$ is not equal to
zero, then 
\[
\xi = \exp(\rmi \, k s / K') \, \xi(0)
\]
with $\xi(0) \ne 0$. Since $\xi$ is an element of $\gothG$, one should have
\[
a'(\xi(0)) = \exp(2 \pi \rmi \, k / K') \, \xi(0).
\]
For any integer $l$ the mapping $\xi'$ defined by
\[
\xi' = \exp(\rmi \, l s) \,\xi
\]
is a nonzero element of $\gothG$. The action of the grading operator $Q$ on
$\xi'$ gives $(K' l + k) \xi'$. The number $K' l + k$ should be an integer.
Since $l$ is an arbitrary integer, it is possible only if $K'$ is an
integer. Actually, due to the remark given after the proof of Proposition
\ref{p:10}, one can assume without any loss of generality that it is a
positive integer.

For any integer $k$ denote by $[k]_{K'}$ the element of the ring
$\bbZ_{K'}$ corresponding to $k$. Let $x$ be an arbitrary element of
$\gothg$ and $\xi$ be an element of $\gothG$ such that $\xi(0) = x$.
Expanding $\xi$ over the grading subspaces,
\[
\xi = \sum_{k \in \bbZ} \xi_k,
\]
one obtains
\[
x = \sum_{m \in \bbZ_{K'}} x_m,
\]
where
\[
x_m = \sum_{\scriptstyle k \in \bbZ \atop \scriptstyle [k]_{K'} = m}
\xi_k(0).
\]
Here for any $m \in \bbZ_{K'}$ we have
\[
a'(x_m) = \exp(2 \pi \rmi \, k / K') \, x_m,
\]
where $k$ is an arbitrary integer such that $[k]_{K'} = m$. Hence, the
automorphism $a'$ is semisimple and $a'^{K'} = \id_\gothg$.

The change $\sigma \to K' \sigma$ induces an isomorphism from $\gothG$ to
$\calL_{a', K'}(\gothg)$ which conjugates the $\bbZ$-gradation of $\gothG$
under consideration with the $\bbZ$-gradation of $\calL_{a', K'}(\gothg)$
generated by grading operator $Q' = - \rmi \, \rmd/\rmd s$. That was to be
proved.
\end{proof}

It follows from the above theorem that to classify all $\bbZ$-gradations of
the twisted loop Lie algebra $\calL_{a,K}(\gothg)$ it suffices to classify
the automorphisms of $\gothg$ of finite order. The solution of the latter
problem can be found, for example in \cite{OniVin90, GorOniVin94}, or in
\cite{Kac94}. Note here that classification of the automorphisms of
$\gothg$ of finite order is equivalent to classification of
$\bbZ_K$-gradations of $\gothg$. Let us have two $\bbZ$-gradations of
$\calL_{a,K}(\gothg)$ which are conjugated to standard $\bbZ$-gradations of
Lie algebras $\calL_{a', K'}(\gothg)$ and $\calL_{a'', K''}(\gothg)$. It is
clear that the initial $\bbZ$-gradations are congugated by an isomorphism
of $\calL_{a,K}(\gothg)$ if and only if $K' = K''$ and the automorphisms
$a'$ and $a''$ are conjugated.

\vskip 1em

{\bf Acknowledgments.} Kh.S.N. is grateful to the Max-Planck-Institut f\"ur 
Gravitationsphysik -- Albert-Einstein-Institut in Potsdam for hospitality 
and friendly atmosphere. His work was supported by the Alexander von 
Humboldt-Stiftung, under a follow-up fellowship program. The work of 
A.V.R. was supported in part by the Russian Foundation for Basic Research 
(Grant No. 04--01--00352).

\appendix

\section{Diffeomorphism groups} \label{a:dif}

Let $M$ and $N$ be two finite dimensional manifolds, and $M$ be compact.
The space $C^\infty(M, N)$ of all smooth mappings from $M$ to $N$ can be
supplied with the structure of a smooth manifold modelled on Fr\'echet
spaces (see, for example, \cite{Ham82, KriMic91, KriMic97}).

Let $K$, $M$, $N$ be three finite dimensional manifolds, and let $M$ be
compact. Consider a smooth mapping $\varphi$ from $K$ to $C^\infty(M, N)$.
This mapping induces a mapping $\slantPhi$ from $K \times M$ to $N$ defined
by the equality 
\[
\slantPhi(p, q) = (\varphi(p))(q).
\]
One can prove that the mapping $\slantPhi$ is smooth. Conversely, if one
has a smooth mapping from $K \times M$ to $N$, reversing the above equality
one can define a mapping from $K$ to $C^\infty(M, N)$, and this mapping is
also smooth. Thus, we have the following canonical identification
\[
C^\infty(K, C^\infty(M, N)) = C^\infty(K \times M, N).
\]
This fact is called the {\em exponential law\/} or the {\em Cartesian
clousedness\/} (see, for example, \cite{KriMic91, KriMic97}).

Let $M$ be a compact finite dimensional manifold. The group $\Diff(M)$ of
smooth diffeomorphisms of $M$ is an open submanifold of the manifold
$C^\infty(M, M)$. Here $\Diff(M)$ is a Lie group. The Lie algebra of
$\Diff(M)$ is the vector space $\Der C^\infty(M)$ of all smooth vector
fields on $M$ equipped with the negative of the usual Lie bracket 
(see, for example, \cite{Mil84, KriMic91, KriMic97}).

Let $\lambda: \bbR \to \Diff(M)$ be a smooth curve through the point
$\id_M$. For each $p \in M$ the curve $\lambda$ induces a curve $\tau \in
\bbR \mapsto (\lambda(\tau))(p)$ in $M$ through $p$. The tangent vector to
this curve at the point $p$ is an element of $T_p(M)$. In this way we
obtain a vector field on $M$ which is the tangent vector to the curve
$\lambda$ at the point $\id_M$. 

Let now $\lambda: \bbR \to \Diff(M)$ be a one-parameter subgroup of
$\Diff(M)$. This means that $\lambda$ is a smooth curve in $\Diff(M)$ which
satisfies the equality
\[
\lambda(0) = \id_M,
\]
and the relation
\[
\lambda(\tau_1) \circ \lambda(\tau_2) = \lambda(\tau_1 + \tau_2).
\]
The mapping $\lambda$ is an element of $C^\infty(\bbR, C^\infty(M, M))$.
Denote the corresponding element of $C^\infty(\bbR \times M, M)$ by
$\slantLambda$. The mapping $\Lambda$ satisfies the equality
\[
\slantLambda(0, p) = p 
\]
and the relation
\[
\slantLambda(\tau_1, \slantLambda(\tau_2, p)) = \slantLambda(\tau_1 +
\tau_2, p).
\]
Hence, the mapping $\slantLambda$ is a flow on $M$. Here the tangent vector
to the curve $\lambda$ at $\id_M$ is the vector field generating the flow
$\slantLambda$.

Since $M$ is a compact manifold, then for each vector field $X$ there is a
flow $\slantLambda^X$ generated by~$X$. This flow induces the one-parameter
subgroup $\lambda^X$ of $\Diff(M)$. It is clear that
\[
\lambda^X(\tau) = \exp (\tau X).
\]
Therefore, in such a way we realize the exponential mapping for $\Diff(M)$.

\section{\mathversion{bold}Distributions on $S^1$ and generalisations}
\label{a:dis}

A continuous linear functional on the Fr\'echet space $C^\infty(S^1) =
C^\infty(S^1, \bbC)$ is said to be a {\em distribution\/} on $S^1$. For a
general presenation of the theory of distributions we refer to the book by
Rudin \cite{Rud73}.

The {\em support\/} of a function $\varphi \in C^\infty(S^1)$ is defined 
as the closure of the set where $\varphi$ does not vanish and denoted as
$\supp \varphi$. We say that a distribution $T$ vanishes on an open set 
$U$ if $T(\varphi) = 0$ whenever $\supp \varphi \subset U$. Then the
support of $T$ is defined as the complement of the union of all open sets
where $T$ vanishes. It is clear that the support of a distribution on $S^1$
is a closed set.

If the support of a distribution $T$ coincides with a one-point set
$\{p\}$, then
\[
T(\varphi) = \sum_{m=0}^n c_m \, \varphi^{(m)}(p).
\]
for some nonnegative integer $n$ and constants $c_m$.

Let now $T$ be a continuous linear mapping from $\calL(\gothg)$ to
$\gothg$. Given a basis $(e_i)$ of $\gothg$, denote by $(\mu^i)$ the
dual basis of $\gothg^*$. For any element $x$ of $\gothg$ one has
\[
x = \sum_i e_i \, \mu^i(x).
\]
Using this equality, one can write
\[
T(\xi) = \sum_i e_i \, \mu^i(T(\xi)).
\]
Representing a general element $\xi$ of $\calL(\gothg)$ as $\sum_j e_j \,
\xi^j$, one obtains
\[
\mu^i (T(\xi)) = \sum_j \mu^i( T(e_j \, \xi^j)).
\]
Introduce a matrix of distributions $(T^i{}_j)$ on $S^1$ defined by the
relation
\[
T^i{}_j(\varphi) = \mu^i(T(e_j \, \varphi)).
\]
Here $\varphi$ is a smooth function on $S^1$. Now one can write
\[
T(\xi) = \sum_{i,j} e_i \, T^i{}_j(\xi^j).
\]
Thus, the matrix $(T^i{}_j)$ completely determines the mapping $T$.

The support $\supp \xi$ of the element $\xi$ of $\calL(\gothg)$ is
defined as the closure of the set where $\xi$ does not take zero
value. Representing $\xi$ as $\sum_i e_i \, \xi^i$ one concludes that
$\supp \xi = \bigcup_i \supp \xi^i$.  We say that a continuous smooth
mapping $T$ from $\calL(\gothg)$ to $\gothg$ vanishes on an open set
$U$ if $T(\xi) = 0$ whenever $\supp \xi \subset U$. The support of $T$
is defined as the complement of the union of all open sets where $T$
vanishes. It is clear that $\supp T = \bigcup_{i,j} \supp T^i{}_j$, where
$(T^i{}_j)$ is the matrix of distributions on $S^1$ which determines
the mapping $T$ for given dual bases $(e_i)$ and $(\mu^i)$ of $\gothg$
and $\gothg^*$ respectively. If the support of $T$ is a one-point set
$\{p\}$, then one can easily demonstrate that
\[
T(\xi) = \sum_{m=0}^n c_m (\xi^{(m)}(p)).
\]
for some nonnegative integer $n$ and endomorphisms $c_m$ of $\gothg$.

Consider now continuous linear mappings from $\calL_a(\gothg)$ to
$\gothg$ for the case when $\gothg$ is a semisimple Lie algebra and $a$ is
an automorphism of $\gothg$ satisfying the relation $a^K = \id_\gothg$ for
some positive integer $K$. In this case $\calL_a(\gothg)$ can be considered
as a subalgebra of $\calL(\gothg)$ formed by the elements $\xi$ satisfying
the condition
\[
\xi(\varepsilon_K p) = a(\xi(p)).
\]
We denote this subalgebra as $\calL_{a, K}(\gothg)$. Define a linear
operator $A$ in $\calL(\gothg)$ acting on an element $\xi$
in accordance with the relation
\[
A \, \xi(p) = a(\xi(\varepsilon_K^{-1} p)).
\]
An element $\xi \in \calL(\gothg)$ belongs to $\calL_{a, K}(\gothg)$ if $A
\, \xi = \xi$.

For an arbitrary element $\xi \in \calL(\gothg)$ the element $\mbar \xi$
defined as
\[
\mbar \xi = \frac{1}{K} \sum_{m=0}^{K-1} A^{-m} \xi
\]
belongs to $\calL_{a, K}(\gothg)$, and one can extend a continuous linear
mapping $T$ from $\calL_{a, K}(\gothg)$ to $\gothg$ to a continuous linear
mapping $\mbar T$ from $\calL(\gothg)$ to $\gothg$ assuming that
\[
\mbar T(\xi) = T(\mbar \xi).
\]
One can easily show that
\[
\mbar T \circ A = \mbar T,
\]
and that
\[
\supp \mbar T = \supp T.
\]

It is clear that if the support of an element $\xi \in \calL_{a,
K}(\gothg)$ contains a point $p \in S^1$, then it contains also the point
$\varepsilon_K p$. Therefore, the support of an element of
$\calL_{a, K}(\gothg)$ is the union of sets of the form
\[
\mbar{\{p\}\!}\, = \{p, \varepsilon_K p, \ldots, \varepsilon_K^{K-1} p \},
\qquad p \in S^1.
\]
The same is true for the support of an arbitrary continuous linear mapping
from $\calL_{a, K}(\gothg)$ to $\gothg$.

Let $T$ be a continuous linear mapping from $\calL_{a, K}(\gothg)$ to
$\gothg$ whose support is $\mbar{\{p\}\!}\,$. The corresponding mapping
$\mbar T$ has the same support and is invariant with respect to the action
of the automorphism $A$. Using these facts one can obtain that
\[
\mbar T(\xi) = \frac{1}{K} \sum_{l=0}^{K-1} \sum_{m=0}^n c_m (a^{-l}
(\xi^{(m)}(\varepsilon_K^l p)))
\]
for some nonnegative integer $n$ and endomorphisms $c_m$ of $\gothg$.
Restricting the mapping $\mbar T$ again to $\calL_{a, K}(\gothg)$ one has
\[
T(\xi) = \sum_{m=0}^n c_m (\xi^{(m)}(p)),
\]
for any $\xi \in \calL_{a, K}(\gothg)$.

\section{Convergence and series in Fr\'echet spaces} \label{a:ser}

A set $\calD$ is said to be {\em directed\/} if it is supplied with a
binary relation $\succcurlyeq$ satisfying the following properties:
\begin{itemize}
\item[(a)] for any element $\alpha \in \calD$ one has $\alpha
\succcurlyeq
\alpha$;
\item[(b)] if $\alpha \succcurlyeq \beta$ and $\beta \succcurlyeq
\gamma$,
then $\alpha \succcurlyeq \gamma$;
\item[(c)] for any two elements $\alpha, \beta \in \calD$, there
exists an
element $\gamma \in \calD$ such that $\gamma \succcurlyeq \alpha$ and
$\gamma
\succcurlyeq \beta$.
\end{itemize}
The relation $\succcurlyeq$ is called a {\em direction\/} in $\calD$.
Below we use the notation $\calD$ for a general directed set. Given a
countable set $S$, we denote by $\calD(S)$ the set of all finite
subspaces
of $S$, considered as a directed set, where 
$\alpha \succcurlyeq \beta$ if
and only if $\alpha \supset \beta$.

A mapping from a directed set $\calD$ to a topological space $X$ is
called a {\em net\/} in $X$. A net $(x_\alpha)_{\alpha \in \calD}$ in 
a topological space $X$ is said to {\em converge\/} to an element 
$x \in X$, or has {\em limit\/} $x$, if for any neighbourhood $U$ 
of $x$ there is an element $A \in \calD$ such that $x_\alpha \in U$ 
for all $\alpha \succcurlyeq A$. Here one also says that the net
$\{x_\alpha\}_{\alpha \in \calD}$ is {\em convergent\/}. 
If $\calD = \bbN$ and $\succcurlyeq$ is the ordinary order 
relation~$\geq$, nets are sequences with the usual definition 
of convergence. 

Let $X$ and $Y$ be topological spaces, and $f$ be a mapping from $X$
to $Y$. The mapping $f$ is continuous if and only if for any net
$(x_\alpha)_{\alpha \in \calD}$ which converges to $x \in X$ the 
net $(f(x_\alpha))_{\alpha \in \cal D}$ converges to $f(x) \in Y$.

Let $X$ be a topological vector space, $I$ be some countable set, and
$(x_i)_{i \in I}$ be a collection of elements of $X$ indexed by $I$.
The symbol $\sum_{i \in I} x_i$ is called a {\em series\/} in $X$. 
Consider a net $(s_\alpha)_{\alpha \in \calD(I)}$, where
\[
s_\alpha = \sum_{i \in \alpha} x_i.
\]
If the net $(s_\alpha)$ converges to an element $s \in X$ we say that the
series $\sum_{i \in I} x_i$ converges {\em unconditionally\/} to $s$ and
write
\[
s = \sum_{i \in I} x_i.
\]
Here the element $s$ is called the {\em sum\/} of the series 
$\sum_{i \in I} x_i$.

The next proposition is a direct generalisation of the corresponding
proposition for series in normed spaces (see, for example, \cite{Die60}).

\begin{proposition}
Let $X$ be a Fr\'echet space whose topology is induced by a countable
collection of seminorms $(\|\cdot\|_m)$, and $\sum_{i \in I} x_i$
be a series in $X$. If for each $m$ the series $\sum_{i \in I} \|x_i\|_m$ 
converges unconditionally, then the series $\sum_{i \in I} x_i$ also 
converges unconditionally.
\end{proposition}

Let $X$ be a topological vector space whose topology is induced by a
countable family of seminorms $(\|\cdot\|)_m$. If a series 
$\sum_{i \in I} x_i$ in $X$ converges unconditionally and for each 
$m$ the series $\sum_{i \in I} \| x_i \|_m$ also converges unconditionally, 
one says that the series $\sum_{i \in I} x_i$ converges {\em absolutely\/}. 
The above proposition says that for complete $X$ unconditional convergence 
of the series $\sum_{i \in I} \| x_i \|_m$ leads to unconditional
convergence
of the series $\sum_{i \in I} x_i$.

For a series whose terms are positive real numbers, unconditional
convergence is equivalent to absolute convergence. Therefore, in this
case it is customary to say simply about convergence. As for the case of a
general series, one sees that absolute convergence, by definition,
implies unconditional convergence, but in accordance with the
Dvoretzky--Rogers theorem \cite{DvRo50}, for infinite dimensional
topological vector spaces there are series which converge unconditionally,
but do not converge absolutely.

The following proposition can be proved along the lines of the proof of the
corresponding proposition for series in normed spaces (see, for example,
\cite{Die60}).

\begin{proposition} \label{p:1}
Let a series $\sum_{i \in I} x_i$ in a Fr\'echet space converge
absolutely. Assume that the set $I$ is represented as the union of a
countable number of nonempty nonintersecting sets $I_j$,
$j \in J$. For any $j \in J$ the series $\sum_{i \in I_j} x_i$
converges
absolutely and the series $\sum_{j \in J} y_j$, where
\[
y_j = \sum_{i \in I_j} x_i,
\]
converges absolutely. Moreover, one has
\[
\sum_{i \in I} x_i = \sum_{j \in J} \left( \sum_{i \in I_j} x_i
\right).
\]
\end{proposition}


\begin{thebibliography}{**}

\bibitem{Sem03}
M. A. Semenov--Tian--Shansky,
{\em Integrable systems and factorization problems\/},
In: Factorization and Integrable Systems, eds. I. Gohberg, N. Manojlovic
and A. Ferreira dos Santos, Birkh\"auser, Boston, 2003, p.~155--218.

\bibitem{LezSav92}
A. N. Leznov and M. V. Saveliev,
{\em Group-theoretical Methods for Integration of Nonlinear Dynamical
Systems\/},
Birkh\"auser, Basel, 1992.

\bibitem{RazSav97a}
A. V. Razumov and M. V. Saveliev,
{\em Lie Algebras, Geometry, and Toda-type Systems\/},
Cambridge University Press, Cambridge, 1997.

\bibitem{RazSav97b}
A.~V.~Razumov and M.~V.~Saveliev,
{\em Multi-dimensional Toda-type systems\/},
Theor. Math. Phys. {\bf 112} (1997) 999--1022
{\tt [arXiv:hep-th/9609031]}.

\bibitem{GorOniVin94}
V. V. Gorbatsevich, A. L. Onishchik and E. B. Vinberg, 
{\em Lie Groups and Lie Algebras, III. Structure of Lie Groups and Lie
Algebras\/}, 
Encyclopaedia of Mathematical Sciences, vol.~41, Springer, Berlin, 1994. 

\bibitem{RazSavZue99}
A. V. Razumov, M. V. Saveliev and A. B. Zuevsky,
{\em Non-abelian Toda equations associated with classical Lie groups\/},
In: Symmetries and Integrable Systems, ed. A. N. Sissakian, JINR, Dubna,
1999, p. 190--203 
{\tt [arXiv:math-ph/9909008]}.

\bibitem{NirRaz03}
Kh. S. Nirov and A. V. Razumov,
{\em On classification of non-abelian Toda systems\/},
In: Geometrical an Topological Ideas in Modern Physics, ed. V. A. Petrov,
Protvino, 2002, p.~213-–221
{\tt [arXiv:nlin.SI/0305023]}. 

\bibitem{Kac94}
V. G. Kac,
{\em Infinite dimensional Lie algebras\/},
Cambridge University Press, Cambridge, 1994.

\bibitem{PreSea86}
A. Pressley and G. Segal,
{\em Loop Groups\/},
Clarendon Press, Oxford, 1986.

\bibitem{Rud73}
W. Rudin,
{\em Functional Analysis\/},
McGraw-Hill, New York, 1973.

\bibitem{OniVin90}
A. L. Onishchik and E. B. Vinberg,
{\em Lie Groups and Algebraic Groups\/},
Springer, Berlin, 1990.

\bibitem{Ham82}
R. Hamilton,
{\em The inverse function theorem of Nash and Moser\/},
Bull. Am. Math. Soc. {\bf 7} (1982) 65--222.

\bibitem{KriMic91}
A. Kriegl and P. Michor,
{\em Aspects of the theory of infinite dimensional manifolds\/},
Diff. Geom. Appl. {\bf 1} (1991) 159--176.

\bibitem{KriMic97}
A. Kriegl and P. Michor,
{\em The Convenient Setting of Global Analysis\/},
Mathematical Surveys and Monographs, vol. 53, American Mathematical
Society, Providence, RI, 1997. 

\bibitem{Mil84}
J. Milnor,
{\em Remarks on infinite-dimensional Lie groups\/},
In: Relativity, Groups and Topology II, eds. B. S. DeWitt and
R. Stora, North-Holland, Amsterdam, 1984, p. 1007--1057.

\bibitem{Die60}
J. Dieudonn\'e,
{\em Foundations of Modern Analysis\/},
Academic Press, New York, 1960.

\bibitem{DvRo50}
A. Dvoretzky and C. A. Rogers, 
{\em Absolute and unconditional convergence in normed spaces\/},
Proc. Nat. Acad. Sci. USA {\bf 36} (1950) 192--197.

\end{thebibliography}
\end{document}